\documentclass[sigconf,10pt]{article}
\usepackage{amsmath} 
\usepackage{graphicx} 
\def\CN{{\mathcal N}}
\begin{document}

\title{BUBBLE-BLUE a multihop private network based on Bluetooth}

\author{Nadjib Achir\\
nadjib.achir@inria.fr\\
Philippe Jacquet\\
philippe.jacquet@inria.fr\\
{Inria Saclay-Ile-de-France}\\
{Saclay}\\
{France}
}

\maketitle
\begin{abstract}
     The aim of the project is to create a kind of “terrestrial STARLINK\texttrademark” network based on users' smartphones.  The BUBBLE-BLUE (BB) project aims to create private Bluetooth bubbles on top of smartphones. In each private bubble, participants will be able to communicate autonomously, without recourse to private operator networks, neither data nor cellular, relying solely on the Bluetooth technology of smartphones. The routing strategy is based on dynamic Connected Dominant Sets (CDS). We present the specific features of a BB network as well as some simulation results on their routing performance.
\end{abstract}
\newtheorem{theorem}{Theorem}
\section{Introduction}
Space-X's Starlink network relies on 6,000 satellites in low-Earth orbit. There are over 7.4 billion smartphones on the planet's surface, which makes a multiplicative factor above one million  which implies that the smartphones are closer to each other than one thousandth the intersatellite range, even closer in densely populated areas.  The BUBBLE-BLUE (BB) project aims to create private Bluetooth bubbles on top of smartphones. In each private bubble, participants will be able to communicate autonomously, without recourse to private operator networks, neither data nor cellular, relying solely on the Bluetooth technology of smartphones.

This is a dual application for military and civilian use. In the military or similar version, messages will be encrypted before being routed to the various recipients.

\subsection{BUBBLE-BLUE aim and usage}
Smartphones are now universally used in the world. Thanks to their internal battery with rapid recharging capability, they are an extremely resilient communication and information tool. So much so that the weakest element in the overall system actually is the cellular network infrastructure.

We list in the following the various circumstances that can prevent access to the cellular network:
\begin{enumerate}
\item In the event of a natural disaster, such as an earthquake or flood, many cellular stations may have been destroyed.
\item In the event of a terrorist panic, the cellular network may be congested or blocked.

\item A private bubble can be invoked in the event of insufficient cellular coverage, for example during large, concentrated music events, or during mountain or water sports trips, or in white zones, or insufficient battery power, as Bluetooth transmissions are low-power and energy-efficient.

\item In the event of armed conflict or police repression by governments, the cellular network can be closely monitored.
\end{enumerate}

A BUBBLE-BLUE network can hold a few dozen participants (the military equivalent of a platoon, ideally up to 50). The following applications will be developed inside private bubbles or at their borders:
\begin{itemize}
\item Exchanging messages between group members
\item Propagating of geo-coordinate information
\item Broadcasting of live photographs
\item Displaying of instant network topology
\item Evaluating of deployment quality with respect to communications
\item Supporting communication with a higher hierarchical level or with authorized related bubbles
\end{itemize}

\subsection{Particularity of Bluetooth bubbles}
The Bluetooth bubble bear some ressemblance with Mobile Ad hoc networks (MANets), in particular with OLSR~\cite{dumbo}. The main distinctive aspect is the fact that the network interface is entirely dedicated to the support of Bluetooth technology. 

The Bluetooth interface is self organizing to create pico networks made of one random master and several slaves. For example a single nodes may be simultaneously member of several surrounding pico-nets. For the rest of the paper we will consider that two nodes are neighbor if they share at least one piconet at Bluetooth level. The switching time between pico-nets being non negligible, BB application will exclude any real time multi-media application such as audio and video.

Furthermore the Bluetooth technology does not support native broadcast transmission such as Wifi (which made it as the preferred support for MANet). It implies that a broadcast transmission of a packet toward all the neighbors of a node will resolve into a multitude of unicast transmissions, one for each neighbor in each piconet.  

The advantage of Bluetooth technology is that it is a very low power technology which confers a stealth property to its transmission thanks to its low energy emission. Furthermore its low power consumption allows a long autonomy even on smartphone battery. 

Although Bluetooth transmissions are difficult to detect, a “mute / silence” function is provided in BB networks for silencing when passing through a sensitive area. On exiting mute mode, the application will request the retrieval of recent messages exchanged by members which have been missed by the muted user. 

\begin{figure}
\includegraphics[width=10cm]{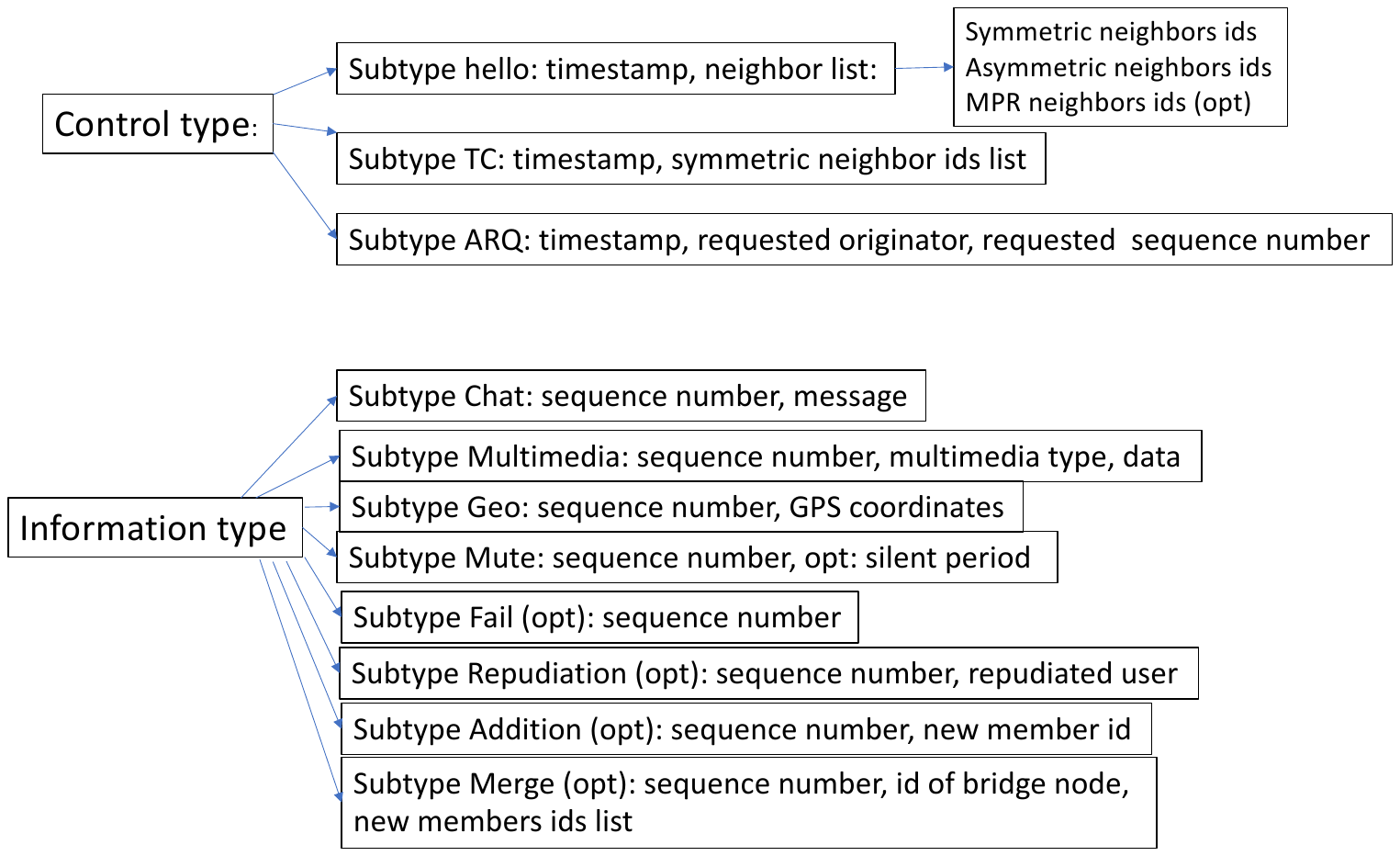}
\caption{The Bubble-Blue function breakdown.}
\label{fig1}\end{figure}
\section{Protocol functioning}
\subsection{Privacy}

We consider a group of $n$ members. Cyber protection of bubbles can be ensured by the prior exchange of symmetrical AES 16 bytes key sets~\cite{aes2}.

For every pair group members $(i,j)$ there is a symmetric key $K_{i,j}$ known by $i$ and $j$.

At mission initialization, the group leader generates an $n\times n$ random matrix $K_{i,j}$ of $n^2$ symmetric 16 bytes keys. The user $j$ should only receive the matrix column $K_{i,j}$ for $i=1,\ldots,n$. This can be done by wired transfer from the leader device.

The application will enable:

\begin{itemize}
\item The tamper-proof exchange of information, well beyond the duration of the mission ({\it e.g.} less than 24 h) thanks to a resilient encryption.

\item The protection of the network and of the protocol integrity. External user will not be capable to affect the network topology control and the routing management

\item {\it A priori}, keys are not renewed for the duration of the mission (less than 24 hours). However, it is possible to repudiate keys if compromise is suspected, and to add members to the network (by direct contact with the group leader), which can be seen as key renewal if it's with a member who has already been repudiated.

\item The interconnection with related bubbles without sharing internal keys by electing gateway members.

\item The possibility of using members of related bubbles as relays between members of the local bubble.

\item Make it possible to create a “relay” app so that local/civil populations can relay private bubbles (also via Bluetooth), and thus extend the mesh.

\item Special military case: the disappearance of the bubble leader. A new leader can be appointed via a pre-arranged chain of command. However, the repudiation function can be delegated. Moreover, the repudiation of the leader will be provided for, for example by consensus among the sub-leaders.
\end{itemize}

\subsection{Packet structure}
The structure of the packet is as follow. For convenience the respective size of the fields are not specified in the following table, see figure~\ref{fig1}

\begin{tabular}{|cccc|c|}
 \hline
 $K_1$&$K_2$&$\ldots$&$K_n$& data\\
 \hline
\end{tabular}

The keys $K_1,\ldots,K_n$ are 16 bytes fields, each specifically dedicated to the members of the bubble. The data field is block encoded by a 16 bytes key $K$ randomly generated by the originator $j$ of the packet: {\it i.e.} data$=C(K,X)$ where $X$ are the original data and $C$ is the AES encoding function. For $i\neq j$ the field $K_i$ is the key $K$ encoded by the key $K_{i,j}$: $K_i=C(K_{ij},K)$. Thus the receiver $i$ must first reverse decode $K_i$ via the symmetric key $K_{ij}$ to get the $K$ and then reverse decode the data field. Any other user, not member of the bubble cannot retrieve the key $K$ and decode the data. The key $K$ is only used for a single packet, while the keys $K_{ij}$ last for the session duration. The particular case $i=j$, key $K_j$ is the field $1,\ldots,1$ since the originator does not need to decode its own data. In passing it indicates the originator of the packet and bootstrap the decoding.

If the member id $i\neq j$ is a repudiated member the field $K_i$ is set to zero, so that the compromised user $i$ cannot decode the data. Similarly, a data generated by a repudiated user will be discarded. 

\subsection{Data field structure}

The decoded Data field structure is of the form

\begin{tabular}{|c|c|c|}
\hline
originator&type&payload\\
\hline
\end{tabular}
The originator is in general the same as the originator determined with the key $K_j=1,\ldots,1$ in the packet header, with the notable exception when the packet contains the data whose retransmission has been triggered by an ARQ packet (and when the user retransmitting the data is not the originator of the data but an intermediate node in the CDS).

The type can be of two kinds: control or information. We first review the control types.

\subsubsection{Control types}

In this case the payload is of the form 

\begin{tabular}{|c|c|c|}
\hline
subtype&timestamp&control data\\
\hline
\end{tabular}

Every member manages a list for each other members with the list of the ten most recent received timestamps.

There should be three control subtypes: hello, topology control (TC), ARQ subtypes. But the latter two subtypes have been moved to information types, since their payload will be carried by packet intended to be flooded via multi-hop broadcasts. 

\paragraph{Subtype Hello}
Purpose: to discover member's neighborhood in the bubble. By neighbor we mean a member which a same bluetooth pico-net with the original member.
In this case the control data is a list of symmetric neighbors followed by a list of asymmetric neighbors. We don't specify the separator field between the two lists, or equivalently the symmetric neighbor list could be headed by a length indicator field. 

A symmetric neighbor of a node $i$ is a neighbor node from which hello packets have been previously and the neighbor node has already received hellos from node $i$. An asymmetric neighbor is a neighbor from which a hello has been received but the link has not been checked in the reverse way, {\it i.e.} the neighbor has not already received a hello from this particular node.

We now review the information types.
\subsubsection{Information  Types}
In this case the payload is of the form

\begin{tabular}{|c|c|c|}
    \hline
    subtype&sequence number&data\\
    \hline
\end{tabular}

Each node maintains a sequence number which is incremented each time an information packet is transmitted.
There are ten subtypes: chat, multimedia TC, arq, geo, mute, fail, repudiation, addition, merge. The two last are optional.

\paragraph{Subtype Chat}
Purpose: to add a message to the group chat. The data is a text. The chat may introduce text limitations in order to make the exchanges more fluid.
\paragraph{Subtype Multimedia}
Purpose to add a multimedia content 9audio, picture video) to the chat stream. Since the bluetooth network constraint may unable to support real time reactive audio video, thus the transmission will be very likely delayed. In some case the data might be split in several pieces, if the whole file exceed the data limit. Each piece will be transmitted with a specific sequence number and will be displayed on a separate window than the chat stream.
\paragraph{Subtype geo}
Purpose: to advertize a GPS position. The data are GPS coordinates and will be displayed on a separate window than the chat stream. When a member is in outdoor position and receive receive a reliable GPS signals, it may broadcast its position. The other members can assume that their position will be the same, since the position differential would not exceed several ten meters or one hundred meters. If two transmissions are broadcasted beyond a too large offset there will be a warning in the chat stream.
\paragraph{Subtype Topology Control}
Purpose: to convey information about the network topology.
The information data is a timestamp and list of members ids. A priori, only the members of the Connected Dominating Set will periodically transmit their symmetric neighbor set. The topology will be displayed on a separate window and be modified upon updates. Optionally this information can be used to compute an efficient covering tree, in the cases wheren (1) the topology is enough stable; (2) the Bluetooth interface allows to select the neighbor to which transmit (instead of selecting all neighbors).

\paragraph{Subtype ARQ}
Purpose to trigger the retransmission of a non received message. The information data is made of a member id plus a sequence number. When a member detects a solution of continuity in the sequence numbers of the information packets it has received from another member, it will trigger the broadcast of an ARQ packet requesting the retransmission of the missing information packet. However the fact of classify the ARQ message as "information" packet may pose problem, because it does not contain any specific information. Furthermore if not received it may trigger further ARQ message just to request a retransmission of the ARQ message with a risk of a infinite chain of ARQ message in case of recurrent failure. Therefore optionally the message could be classified as a control message and would not be subject to subsequent ARQ messages requesting retransmission in case of failure. This option will be left as further work, since it will require a control message to be broadcasted beyond one hop.
\paragraph{Subtype mute}
Purpose: to inform about switching into muted status. There is no data. Optionally the data can contain a time offset during which the node will remain silent. If the node does not demute before this declared muted time it might trigger an alarm and make the group leader to declare  repudiated the late demuted user. The time offset can be declared undefinite if the user does not have a control on its muted time. 
\paragraph{Subtype fail}
Purpose: to inform about password failure. There is no data. When exiting the mute state the bubble application will request the user to insert a password. If the user fails within a certain predefined time period the application will transmit a fail message and trigger the self stopping and data destruction. 
\paragraph{Subtype repudiation}
To advertize that a member is repudiated (only by privilege of group leader). The data is the id of the repudiated member.
Consequence: the public key of the repudiated member is deactivated in every other members (will always be set at zero in their packets). The messages sent by the repudiated member will be subsequently ignored
\paragraph{Subtype addition (optional)}
Purpose: to add a new member to the group. Data: the value $j$ of the new member. The addition is made possible if the columns of the matrix $K_{i,j}$ which have been downloaded by group members prior to the mission already shows provisions for new members ({\it i.e.} extend beyond the current size of the group. Separately the new member must download their $K_{ij}$ column from the group leader in a local transmission, preferably via wires.

\paragraph{Subtype merge (optional)}
Purpose : to enable the fusion of two groups. Data: the value of the id of the bridge node between the two groups plus the list of the ids of the new members in the old group 

\subsection{Routing algorithm}

The nodes discard any packet received beyond a tolerance period around its timestamp. 

Every post keeps and manages a list of symmetric neighbor (S) and of asymmetric neighbors (A) with for each neighbor a date of last reception. 
\subsubsection{Transmitting hellos}
At the end of each inter-hello period the node transmits a hello packet containing the list of its neighbor in S and A status toward all its bluetooth neighbors n(managed in the bluetooth interface via various pico-network). When a node receives a hello packet, it puts the id of the hello generator in the A list. If the node reads its own id in the list of neighbors in the hello packets, it moves the id of the hello originator in the node S list.

If the node does not receive a hello packet from a given other node after a tolerance period equal to an integer multiple of the inter-hello period, it removes this neighbor from its A and S neighbor lists.

Upon reception of an hello packet from a S neighbor the node stores the list of S neighbor listed in the originator hello and create or updates a two hop neighbor table for this particular node.

\subsubsection{CDS Election}
There are various CDS election algorithm, each of them demanding more information on the local topology. 

\paragraph{Wu-Li 1999 algorithm~\cite{wu1999}}
It only requires each node to know the two hop neighborhood, {\it i.e.} to know the neighborhood of each neighbor. This is given by storing the symmetric neighbor list in each hello received from a symmetric neighbor. On node $X$ we denote $\CN(X)$ as the neighbor set. If $X$ id is smaller than the ids of all its symmetric neighbor then $CDS(X)$ is set at true ({\it i.e.} $X$ is member of the CDS). Otherwise if the set of $X$ symmetric neighbors with id {\bf smaller} than X id forms a CDS of the set $\CN(X)$ (checked via the two hop neighbor table), then $CDS(X)$ is false, otherwise it is set at true. 

\paragraph{MPR CDS ~\cite{jacquet2004}} The MPR CDS requires that each node $X$ selects an MPR subset of $\CN(X)$. This would need that each node indicates in its hello packet which symmetric neighbors is member of the MPR set. This adds a third list in the hello packet dedicated to the MPR set. We don't describe the MPR set selection which has already has triggered a significant literature. A node $X$ is member of the MPR CDS iff either $X$ id is smaller than the ids of all its symmetric neighbors or it is MPR of the neighbor with the smallest Id when its id is smaller than X's id.  The MPR CDS is in general smaller than the CDS set obtained with the 1999 Wu-Li algorithm but it requires more information.

\paragraph{The MPR CDS flooding set~\cite{jacquet2004}}
The flooding via the MPR set, which is the foundation of the routing protocol OLSR, is very performant but requires more local information. It is based on the MPRs flooding, and requires that each information packet received by a node is delivered with the id of the last transmitters, which needs an adatation of the Bluetooth layer. Furthermore, since each broadcast resolves in several unicast successive transmission, the reception order of packets copies may differ from one node to another node and may lead to MPR flooding malfunction. This algorithm is not advised for Blue-Bubble.  

\paragraph{Wu-Li 2001 algorithm~\cite{wu2001}}
This algorithm is more complicated than the first algorithm in~\cite{wu1999}. It operates on marked and unmarked nodes and necessitates several rounds where the marked status is advertized by the nodes. It is therefore more prone to errors in case of synchronisation glinches and in case of out of order receptions. It seems to give performance close to optimal in 1D.

\paragraph{Optimal CDS selection}
It should be noted that the optimization of the CDS should not be solely based on the size of the CDS, but should also include the internal and external degrees of the CDS, since any broadcast over Bluetooth resolve into a multitude of unicast transmissions. The optimal CDS search is an NP complete problem but the limited size of the Blue-Bubbles may make feasible. However it needs a complete knowledge of the network topology which would require that each node in the network transmit a TC advertizing its own full neighbor list. This poses an egg and chicken problem because the knowledge of the full topology would need an efficient broadcast 

\paragraph{Covering tree flooding}
The optimal broadcast will be attained with the covering tree. Indeed the broadcast via bluetooth does not provide any saving since it resolves in all cases in successive unicast transmissions. There are many covering tree algorithms, in all cases the cost of flooding is $n-1$ transmission, $n$ being the size of the BUBBLE-BLUE network. The determination of the covering tree requires the knowledge of the full topology, as for the optimal CDS. Furthermore the operations of the covering tree require that the interface between the BUBBLE-BLUE application and Bluetooth, allows a selection of the destination of the local transmissions of the copies of the packet. This is mandatory in order to avoid wasteful transmissions outside the covering tree.

\subsubsection{Broadcast transmissions}
A node which desires to broadcast an information, increments its sequence number and sends the packets toward all its neighbors. Only the neighbors which have not already seen the sequence number will process the packet, otherwise the packet is discarded. Only the members of the CDS ({\it i.e.} the nodes $X$ with $CDS(X)$ at true) retransmits the packet toward all their neighbors {\it without any change}, if not discarded except if the sequence number is smaller than the sequence numbers stored about the originator in the CDS member sequence list.


\paragraph{Topology broadcast} 
Each post maintain a topological directory which will be the list of the members of the CDS with a timestamp and their list of S neighbor, together with their timestamp. 

Periodically the CDS members broadcasts a TC packet with its list of S neighbors with their time stamp. Upon the reception of a TC packet with generator $Y$ not seen before, each user updates their topological directory wrt $Y$, only the entries with timestamp smaller than the timestamp of the entry. If the user is member of the CDS its retransmit the packet toward all its neighbors.

If a topological directory is not updated within a period multiple of update periods then the entry is discarded.
\paragraph{ARQ broadcast}
Each user maintain a list of sequence numbers for each other user. 
Upon reception of an ARQ packet by a non CDS member, if the requested sequence number is in the sequence number list received for the intended user which is the originator of the missing packet in the information field, then the ARQ receiver does nothing. If the user is member of the CDS and the requested sequence is not in its sequence list then it retransmit the ARQ packet. If the CDS member has the missing packet, it retransmits it in behalf of the missing packet generator and does not retransmit the ARQ packet. In this case the originator of the retransmitted packet will not be the originator of the original paper. It is one of the special case where the actual originator of the packet (with the key in the header equal to $1,\ldots,1$) is not the advertized originator in the payload. If on the contrary the receiver of the ARQ packet is the actual generator of the missing packet, it retransmits the requested packet. 

Optionally (not described in this paper) the ARQ packet could be turned as a control packet, witht the unique characteristic that it may be routed in a multi-hop broadcast retransmission. As a control packet it woult not trigger subsequent ARQ or ARQ retransmissions in case of failure.

\section{Models and Simulations}
Let us define a graph $G = (V, E)$ with a set of nodes $V = \{1, \dots, n\}$ and a set of edges $E$. Let $u\in V$ we denote $\deg(u)$ the degree of $u$ in $G$. Let $V'\subset V$ by extension we denote $\deg(V')$ the sum of the degrees of all vertices in $V'$.

\begin{theorem}
Assuming $V'$ forms a CDS the average cost of flodding in $G$ via $V'$ is equal to $2\frac{|E|}{n}+(1-\frac{1}{n})\deg(V')$.
\end{theorem}
\paragraph*{proof}
If the initiator $u$ of the broadcast transmission is in $V'$ then the broadcast costs $\deg(V')$ transmissions. If $u\notin V'$ then the broadcast costs $\deg(u)+\deg(V')$ transmissions. The average broadcast cost is $\frac{1}{n}\sum_{u\notin V'}\deg(u)+\deg(V')$. Since $\sum_{u\notin V'}\deg(u)=\sum_{u\in V}\deg(u)-\deg(V')$, and giving the identity $\sum_{u\in V}\deg(u)=2|E|$, it yields the desired result.

If the interface with Bluetooth allows to make a "check valve", {\it i.e.} that the interface provides the identity of the last transmitter in order to avoid to transmit back the same packet, then the quantity $\deg(V')$ should be halved. However the check valve is a specific complication of the application Blue-Bubble.

We will model the graph as an unit disk graph in dimension 1 and dimension 2. In dimension 1 (resp. 2) the nodes are dispatched in a one unit segment (resp $1\times 1$ square) or in a segment of length $\ell$ (resp. $1\times\ell$ rectangle), with the assumption that a two nodes $u$ and $v$ form an edge iff the euclidean distance between $u$ and $v$ is smaller than the unity. In the following we denote by $\lambda$ the node density. The average neighbor size in dimension 1 is $2\lambda$ and in dimension 2, it is $\pi\lambda$.
\subsection{1D Unit disk graph}
We consider a 1D unit disk graph with a density $\lambda$. The results in~\cite{qayyum2002} and~\cite{jacquet2004} show that when $\lambda\to\infty$
\begin{itemize}
    \item The average density of the MPR CDS flooding set is $1$.
    \item The average density of the MPR CDS is $\frac{3}{2}$.
    \item The average density of the 1999 Wu-Li CDS set is $2$
\end{itemize}
Consequently, assuming that the 1D network spans on a segment of length $\ell>1$
\begin{theorem}
The average degree of the MPR flooding CDS set is $2\ell\lambda$; the average degree of the MPR CDS set is $3\ell \lambda$; the average degree of the Wu-Li CDS set is $4\ell\lambda$.
\end{theorem}
Consequently, the average broadcast cost of the MPR flooding set is where $n$ is the number of members of the Blue-Bubble network $2\lambda+2\ell\lambda(1-1/n)=2(\ell+1)\lambda-2$ since $n=\lambda\ell$. The average degree of the MPR CDS set is $(3\ell+2)\lambda-3$, and for the Wu-Li CDS set it is $(4\ell+2)\lambda-4$. 

If a "check valve" is installed, then these figures will now be $(\ell+2)\lambda-1$, $(\frac{3}{2}\ell+1)\lambda-3/2$, and $2(\ell+1)-2$. In comparison, the covering tree is $\lambda\ell-1$, but its application in the Blue-Bubble would need more than a check valve.

The figures~\label{figcs1} and~\label{figds2} confirm the analytic results for the optimal CDS, the 1999 Wu-Li algorithm and the MPR CDS set algorithm. They have been obtained by simulation on a network segment of length 10. 

\begin{figure}[htp]
  \centering \includegraphics[width=\linewidth]{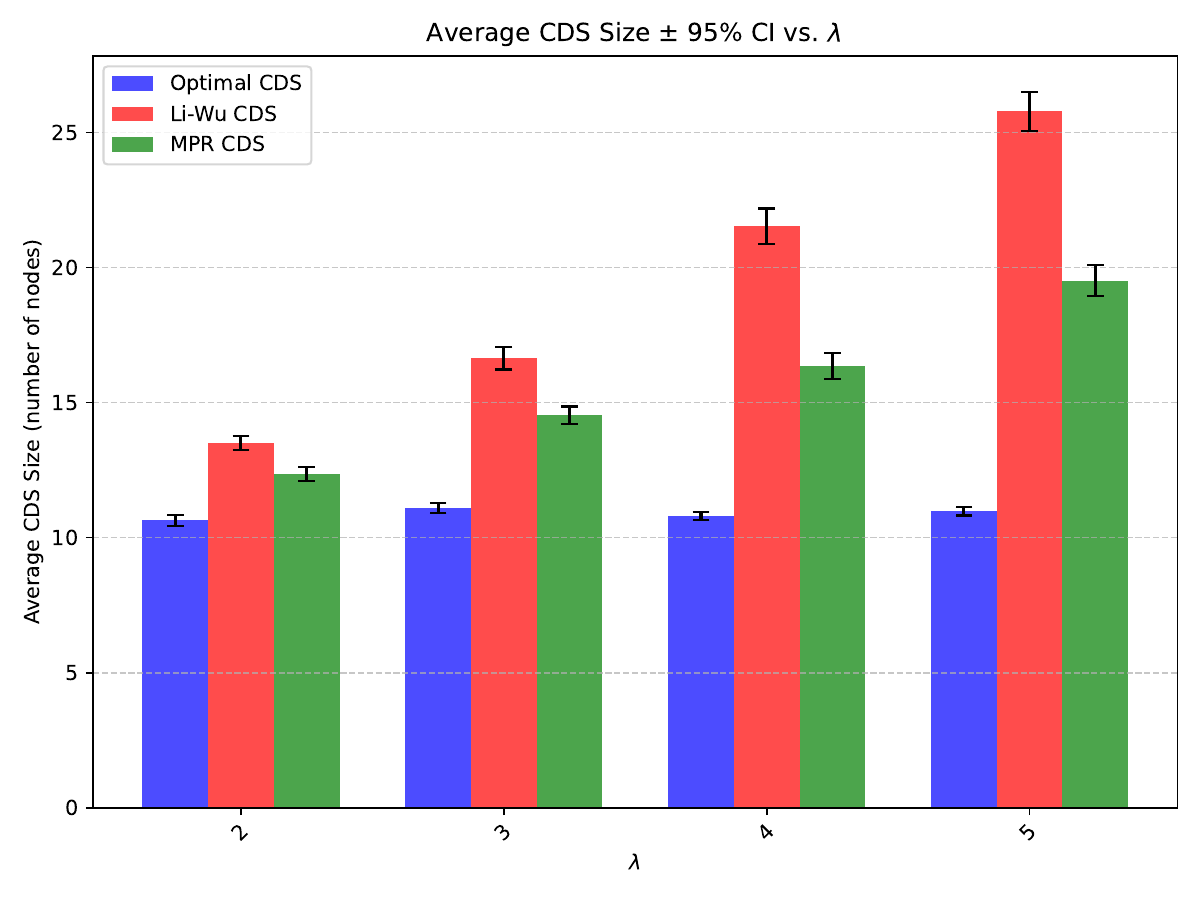}
  \caption{Average CDS size simulated in 1D unit disk graph model.}
\label{figcds1}\end{figure}

\begin{figure}[htp]
  \centering \includegraphics[width=\linewidth]{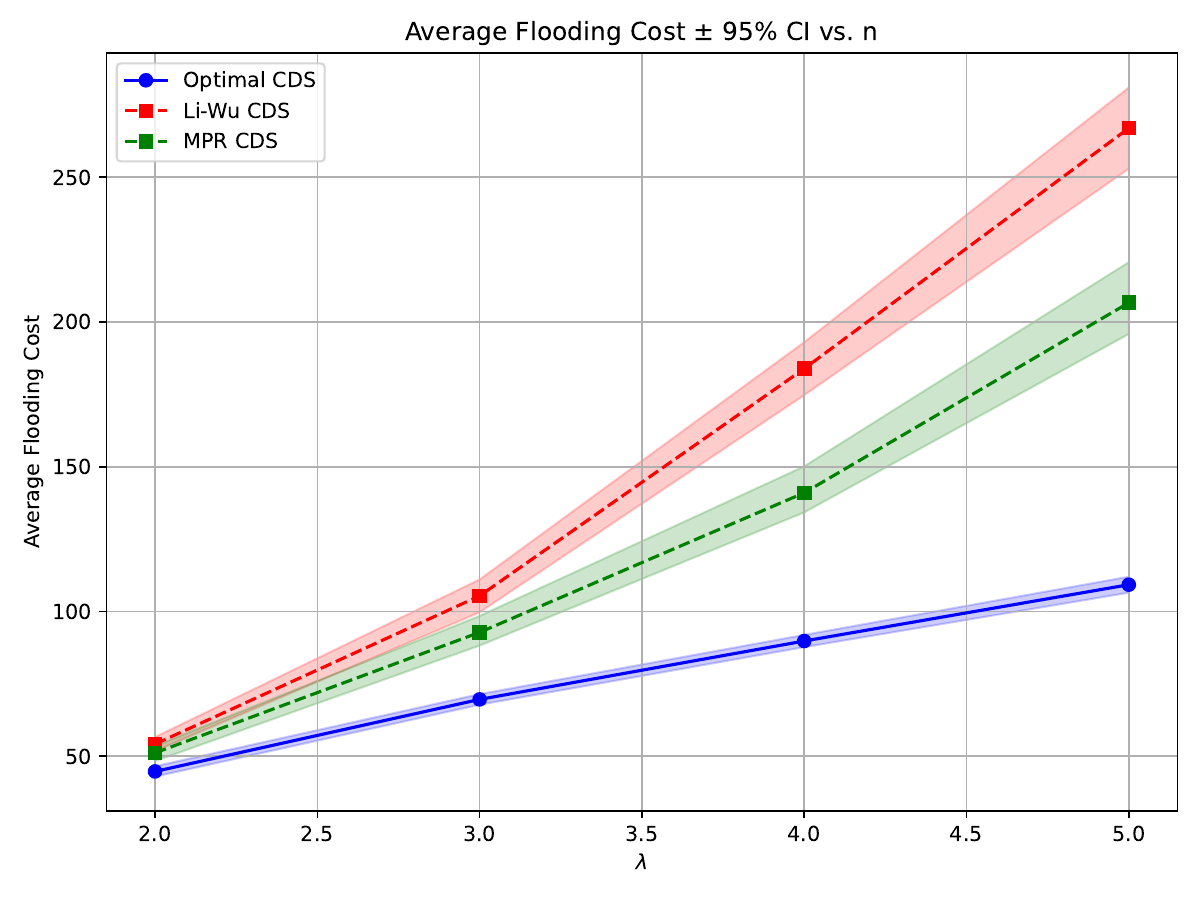}
  \caption{Average CDS size  simulated in 2D unit disk graph model.}
\label{figcds2}\end{figure}



\subsection{2D unit disk graph, optimal CDS code}
The figures~\label{figcs3} and~\label{figds4} confirm the analytic results for the optimal CDS, the 1999 Wu-Li algorithm and the MPR CDS set algorithm. They have been obtained by simulation on a unit disk graph network on a square area of side length 1. The algorithm Wu-Li performs better than the MPR set. Its simplicity makes it our favorite for Blue-Bubble future implementation.

For both set of simulations we have simulated the search for the optimal CDS,  {\it i.e.}, finding the smallest possible subset $D \subseteq V$ such that every node in the graph is either included in this subset or is directly adjacent to at least one node in it. To model this, we introduce binary variables $x_i$ for each node $i$. This variable equals $1$ if node i is included in the dominating set and $0$ otherwise.

The goal is to minimize not simply the cardinality of the dominating set but rather the sum of the degrees of the selected nodes so that the objective function explicitly accounts for the structural cost associated with high-degree nodes. Formally, the objective is formulated as $\min \sum_{i \in V} \deg(i) \, x_i$. To ensure that every node is covered, we impose the constraint that for each node $i$, either $i$ itself must be in the dominating set or at least one of its neighbors must be; formally, this is expressed by $x_i + \sum_{j \in \CN(i)} x_j \geq 1$ where $\CN(i)$ denotes the set of neighbors of node $i$.

In addition to selecting the nodes, we must ensure that the chosen nodes create a single connected component. To achieve this, the model includes continuous flow variables $f_{ij} \geq 0$ for each directed edge $(i, j)$. These flow variables are used in a single-commodity flow formulation that guarantees connectivity: all flow must be routed through the selected nodes only, forming one connected structure rooted at a designated root node $r$. 

Formally, the root node must send out exactly $k-1$ units of flow, each selected non-root node must receive exactly one unit of flow, and nodes that are not in the dominating set neither send nor receive flow. These conditions are described by the system of equations

\begin{equation}
\sum_{j : (j,i) \in E} f_{ji} - \sum_{j : (i,j) \in E} f_{ij} =
\begin{cases}
k - 1, & i = r \\[6pt]
-1, & i \neq r,\; x_i = 1 \\[6pt]
0, & x_i = 0
\end{cases}
\quad \forall i \in V.
\end{equation}

Additionally, to prevent flow from passing through edges that are not incident to selected nodes, we impose the capacity constraints $f_{ij} \leq (|V|-1) x_i$ and $f_{ij} \leq (|V|-1) x_j$ for all edges $(i,j)$. Finally, the root node must always be included in the dominating set, so $x_r = 1$. 

The complete optimization problem combines all of these elements and can be summarized as follows: 
$\min_{x,f} \sum_{i \in V} \deg(i) \, x_i$ subject to

\begin{equation}
\left\{
\begin{array}{ll}
\forall i \in V, &x_i + \sum_{j \in \CN(i)} x_j \geq 1, \\
i = r,\;&\sum_{j : (j,i) \in E} f_{ji} - \sum_{j : (i,j) \in E} f_{ij} = k - 1  \\
i \in V \setminus \{r\},  x_i = 1  & \sum_{j : (j,i) \in E} f_{ji} - \sum_{j : (i,j) \in E} f_{ij} = -1 \\
 i \in V,\; x_i = 0& \sum_{j : (j,i) \in E} f_{ji} - \sum_{j : (i,j) \in E} f_{ij} = 0  \\
\forall (i,j) \in E & f_{ij} \leq (|V|-1) x_i   \\
\forall (i,j) \in E & f_{ij} \leq (|V|-1) x_j  \\
 x_r = 1 
\forall i \in V & x_i \in \{0,1\}   \\
\forall (i,j) \in E & f_{ij} \geq 0 .
\end{array}
\right.
\end{equation}

\begin{figure}[htp]
\centering
    \includegraphics[width=\linewidth]{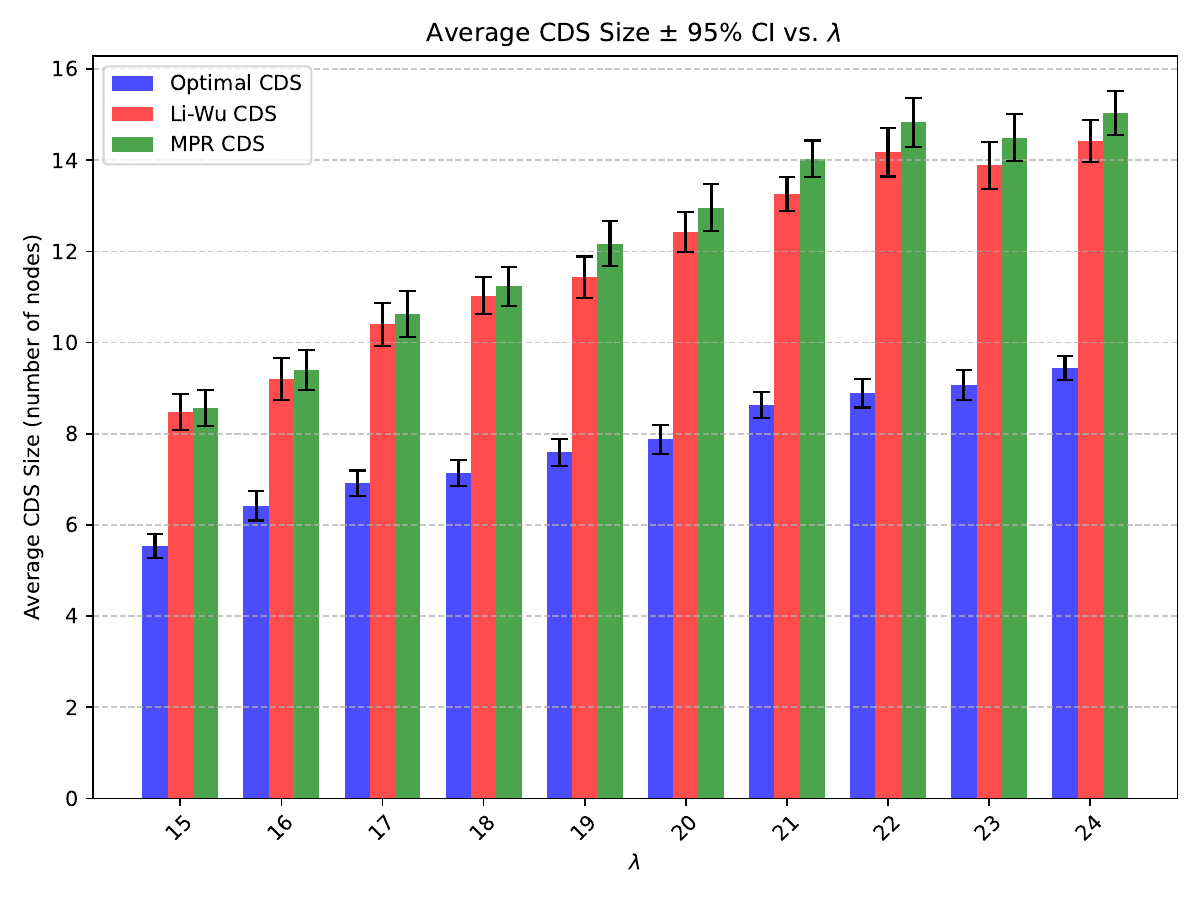}
  \caption{Average CDS size simulated in 2D unit disk graph model.}
\label{figcds3}\end{figure}

\begin{figure}[htp]
  \centering \includegraphics[width=\linewidth]{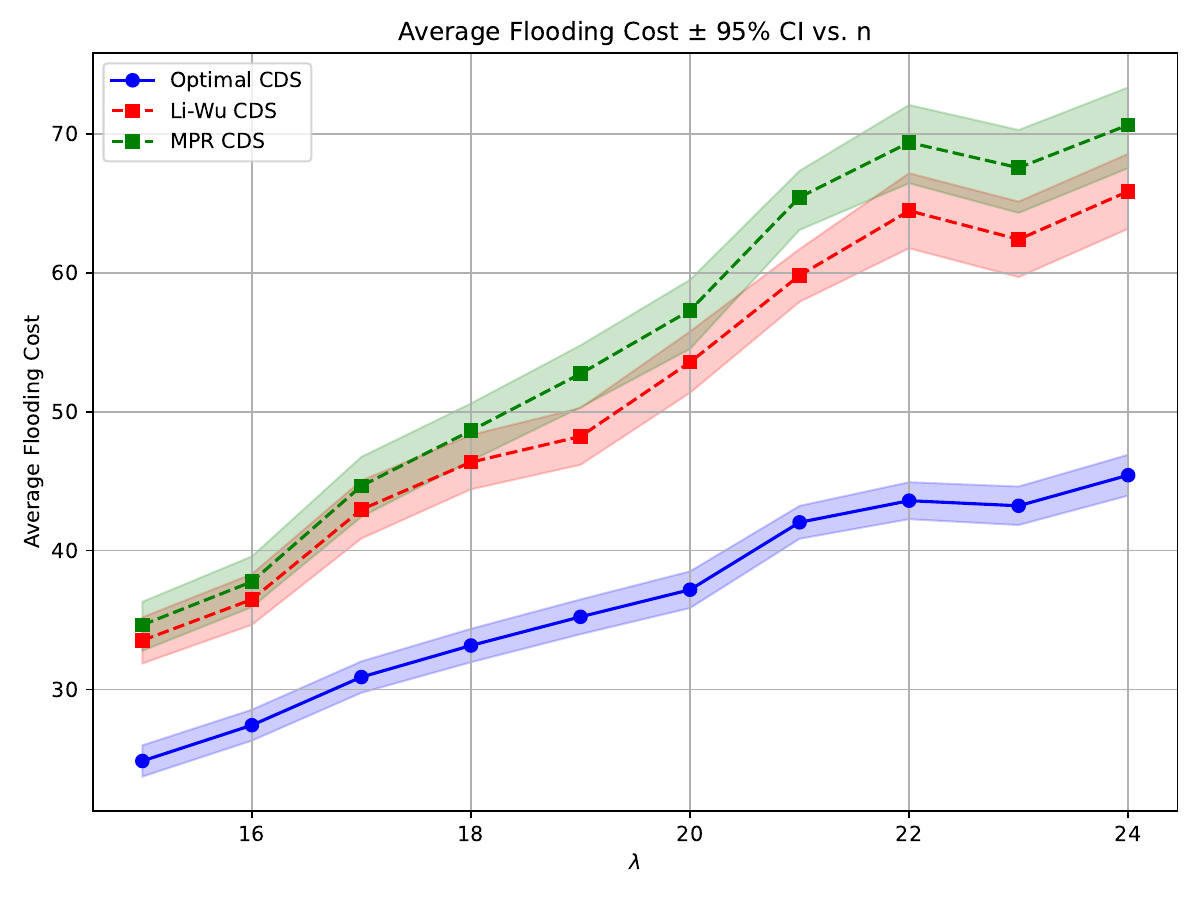}
  \caption{Average CDS flooding cost simulated in 2D unit disk graph model.}
\label{figcds4}\end{figure}

\section{Conclusion}
We have presented a new model of mobile local area network, BUBBLE-BLUE. It is based on smartphone inborn Bluetooth package in order to insure stealthiness and autonomy. BB has dual purposes, military and civilian. The military use implies a high level security based on Advanced encryption standard (AES). In both usage the backbone of the communication is a dynamically elected Connected Dominating Set (CDS). In this paper we analyses the performances and drawbacks of the various CDS election algorithm which are applicable to BB. Contrary to WiFi based Mobile {\it ad hoc} networks, the Bluetooth based BB implies that each local broadcast transmission resolve in many parallel unicast transmission, sometimes with redundancies, at pico-net level, and this affect the performance of the network, and hamper multimedia real time usage.

\section{Acknowledgement}
The authors thanks the society OLVID in the person of Matthieu Finiasz and Lunabee in the person of Thomas Jaussouin and Olivier Berni for their expertize and help in security and cryptography and for their support in the Blue-Bubble project building.



\bibliographystyle{ACM-Reference-Format}
\bibliography{arxivBB}


\end{document}